# Accurate Ground State Electronic and Related Properties of Hexagonal Boron Nitride (h-BN)

Y. Malozovsky, C. Bamba, A. Stewart, L. Franklin, and D. Bagayoko
Department of Physics, Southern University and A&M College, Baton Rouge, LA 70813

## Abstract

We present an ab-initio, self – consistent density functional theory (DFT) description of ground state electronic and related properties of hexagonal boron nitride (h-BN). We used a local density approximation (LDA) potential and the linear combination of atomic orbitals (LCAO) formalism. We rigorously implemented the Bagayoko, Zhao, and Williams (BZW) method, as enhanced by Ekuma and Franklin (BZW-EF). The method ensures a generalized minimization of the energy that is far beyond what can be obtained with self-consistency iterations using a single basis set. The method leads to the ground state of the material, in a verifiable manner, without employing over-complete basis sets. Consequently, our results possess the full, physical content of DFT, as per the second DFT theorem [AIP Advances, 4, 127104 (2014)]. We report the ground state band structure, band gap, total and partial densities of states, and electron and hole effective masses. Our calculated, indirect band gap of 4.37 eV, obtained with room temperature experimental lattice constants of a = 2.504 Å and c = 6.661 Å, is in agreement with the measured value of 4.3 eV. The valence band maximum is slightly to the left of the K point, while the conduction band minimum is at the M point. Our calculated total width of the valence and total and partial densities of states are in agreement with corresponding, experimental findings.

PACS: 71.15.Mb, 71.20.Mq. 71.20.Nr, 71.18.+y

## 1. Introduction

The demand for compact ultraviolet laser devices has led many researchers to search for materials with band gaps larger than that of GaN (3.4 eV), a material presently utilized in the fabrication of high-power, blue-ray laser devices [1]. Properties of hexagonal boron nitride (h-BN), with a graphite-like crystal structure, provide a basis for many applications. It is employed as a good electrical insulator, with excellent thermal conductivity, for crystal growth and molecular beam epitaxy. It has several applications in electronics and nuclear energy industries and serves as an excellent lubricant [2]. Recently, its outstanding catalyst properties have attracted much attention, for potential applications in oxygen reduction reactions [3-5]. Hexagonal boron nitride (h-BN) is a wide band gap material with high chemical and thermal stability. Despite the above attributes of h-BN, a survey of the literature shows a lack of consensus on the experimentally determined band gap of the material. Measured, direct and indirect band gaps have been reported, with values ranging from 3.6 to 7.1 eV. Its electronic structure and band gap have been studied experimentally using x-ray photoemission [6 - 9], optical absorption [10], UV absorption [11], optical reflectivity [12, 13], luminescence spectra [14, 15], photoconductivity [16, 17], and temperature dependence of the electrical resistivity [18]. The various experimentally measured band gaps are summarized in Table 1. From the content of the table, we infer a lack of consensus not only on the direct or indirect nature of the band gap, but also on its numerical value – notwithstanding some of the discrepancies may be due to differences in sample purity, thickness (for films) and measurement temperature.

**Table 1.** Experimental values of the band gap ($E_g$) of h-BN, in eV. The results in this table are reportedly for bulk h-BN. We note that some authors believe the measured indirect band gap of 4.3 eV [9-11] best represents the true band gap of h-BN.

| Experimental method | $E_g$ (eV) |
|---|---|
| X-ray photoemission spectra | 3.6 [a], 3.85 [b] |
| Optical and UV absorption | 3.9 [c], 4.3 [d] |
| Laser – induced fluorescence (LIF) | 4.02 [e] |
| Optical reflectivity spectra | 4.5 [f], 5.2 [g] |
| Luminescence optical spectra | >5.5 [h], 5.89 [i], 5.95[j] |
| Photoconductivity, and absorption spectra | 5.8 [k], 5.83 [l] |
| Temperature dependence of electrical resistivity | 7.1 [m] |

[a] Ref. [6], [b]Ref. [7-9], [c]Ref. [10], [d]Ref. [11-13], [e]Ref. [14], [f]Ref. [15], [g]Ref. [16], [h]Ref. [17], [i]Ref. [18], [j]Ref. [19], ], [k]Ref. [20], [l]Ref. [21], [m]Ref. [22]

As shown in Table 2, the theoretical studies of h-BN disagree on the value of the band gap and particularly on the locations of the valence band maximum (VBM) and of the conduction band minimum (CBM), respectively. Specifically, the table shows that previous LDA and GGA calculations [22-32] led to seven (7) different pairs of VBM and CBM: M-H (1), H-M (5), K-M (2), M-K (1), H-K (1), Γ–H (1) and Γ–K (2), where the numbers between parentheses represent the respective frequencies of the concerned VBM-CBM pair. The two Green function and dressed Coulomb approximation (GW) calculations in the table found the gap to be from H to M. With an LDA potential, Ma et al. [23] employed the linear combination of pseudo-atomic-orbitals (PAO) method to calculate properties of h-BN. Their calculated, indirect band gap, from H to M, was 3.7 eV [23]. The calculated direct (H-H) and indirect (H-M) band gaps, obtained by using the Full Potential Linearized Augmented Plane Wave (FP-LAPW) method, were respectively 4.3 eV and 3.9 eV [24]. The LDA pseudopotential calculations of Blasé et al. [25] resulted in an indirect (K-M) band gap of 3.9 eV while their GW quasiparticle calculations produced an indirect (H-M) band gap of 5.4 eV. Xu and Ching [26], using orthogonalized linear combination of atomic orbitals (OLCAO), found an indirect (K-M) band gap of 4.07 eV. The optimized ultra-soft (Vanderbilt-type) LDA pseudopotential calculations of Furthmüller et al. [27] predicted an indirect (H-M) band gap of 4.1 eV and a direct (M-M) gap of 4.5 eV. Table 2 shows the above referenced results and several other theoretical findings [28-33].

**Table 2.** Illustrative, previously calculated values of the band gap ($E_g$) of h-BN, in eV. They include results from LDA, GGA, and GW calculations.

| Computational method | Potentials | $E_g$ (eV) |
|---|---|---|
| Linear Combination of Pseudoatomic Orbitals (LCPAO) | LDA | 3.7 (M-H) [a] |
| FP-LAPW | LDA | 3.9 (H-M) [b]<br>4.3 (H-H) [b] |
| Ab-initio pseudopotential | LDA | 3.9 (K-M) [c] |
| OLCAO | LDA | 4.07 (M-K) [d] |
| Ultra soft Pseudopotential | LDA | 4.1 (H-M) [e]<br>4.5 (M-M) [e] |
| FP-LAPW | LDA | 4.0 (H-M) [f]<br>4.5 (M-M) [f] |
| FP-LAPW | LDA | 4.58 (H-K) [g] |
| FP-LAPW | PW91-GGA | 4.53 (Γ-K) [g] |
| FP-LAPW | PBE-GGA | 4.54 (Γ-K) [g] |
| Projected-Augmented-Wave (PAW) | LDA | 4.02 (K-M) [h] |
| PAW (VASP) | LDA | 4.21 (H-M) [i] |
| PAW (VASP) | GGA | 4.39 (H-M) [i] |
| PAW | GGA | 4.47 (K-M) [j] |
| GW | GGA | 5.4 (H-M) [c] |
| GW | LDA | 5.95 (K-M) [h] |
| GW | LDA | 5.95 (H-M) [k] |

[a]Ref. [23] [b]Ref. [24], [c]Ref. [25], [d]Ref. [26], [e]Ref. [27], [f]Ref. [28], [g]Ref. [29], [h]Ref. [30], [i]Ref. [31], [j]Ref. [32], [k]Ref. [33].

Clearly, this range of theoretical results for the band gap of h-BN, including the seven (7) different pairs of VBM-CBM, points to the need for further work. Additionally, and unlike the cases for most semiconductors, the experimental results in Table 1 also disagree. These discrepancies constitute a major motivation for this work. This motivation is partly predicated on previous, theoretical results of our group, in agreement with corresponding experimental ones, for more than 30 semiconductors [34].

## 2. Method and Computational Details

We succinctly provide below the essential features of our computational approach. Extensive details on it are available in the literature [34-41]. As with most other calculations, we employed a density functional theory (DFT) potential and the linear combination of atomic orbitals (LCAO). Our specific DFT potential for this work is the local density approximation (LDA) one by Ceperley and Alder, with the parameterization of Vosko et al. [42-45]. A major difference between our method and most others in the literature stems from our performance of a generalized minimization of the energy functional to attain the ground state of the system, without utilizing over-complete basis sets. The first [46 - 48] and the enhanced [49 - 51] versions of this generalized minimization of the energy are respectively expounded upon in the literature.

As per the second DFT theorem, self-consistent iterations with a single basis set lead to a stationary solution among an infinite number of such solutions. This fact resides in the reality that the ground state charge density (i.e., basis set) is not *à priori* known, as far as we can determine. Consequently, the chances are extremely small for a calculation with a single basis set to lead to the ground state of the system or to avoid over-complete basis sets.

We have described in previous publications a straightforward way to search for and to reach the ground state of the system. Beginning with a small basis set that is large enough to account for all the electrons in the system, we perform successive self-consistent calculations, where the basis set of a calculation, except for the first one, is that of the preceding calculation augmented with one orbital. The first and second versions of our method, known as BZW and BZW-EF method, differ as follows. For the first one, we add orbitals in the order of increasing energy of the excited states they represent. In the second, we heed the "arbitrary variations" clause of the second DFT theorem and add orbitals so as to recognize the primacy of polarization orbitals (p, d, and f) over the spherical symmetry of s orbitals for valence electrons. Indeed, for diatomic and any other multi-atomic system, valence electrons do not possess any full, spherical symmetry known to us, unlike the core electrons. The above referenced, successive calculations continue until three (3) consecutive ones produce the same occupied energies. This criterion guarantees the attainment of the absolute minima of the occupied energies (i.e., the true ground state). With just two (2) consecutive calculations leading to the same occupied energies, these energies could represent a local minima and not the absolute ones. The first of the referenced three (3) consecutive calculations [34] is the one providing the DFT description of the material. The basis set for this calculation is dubbed the *optimal basis set*, i.e., the smallest basis set leading to the ground state charge density and energies.

In this study, we utilized the program package developed at the US Department of Energy's Ames Laboratory, in Ames, Iowa. B and N are light enough to neglect relativistic corrections. Self-consistent calculations of the electronic energies and wave functions for the atomic or ionic species provided input data for the solid-state calculations. Specifically, for hexagonal BN, the species we considered were $B^{3+}$ and $N^{3-}$. Preliminary calculations for neutral atoms (B and N) pointed to a charge transfer larger than 2, from B to N.

We provide below computational details to enable the replication of our work. Hexagonal BN (h-BN) belongs to the $D_{6h}^4$ space group, with a space group number of 194, a Pearson symbol of hP4, and Patterson space group P6$_3$/mmc [17]. There are two atoms of each kind in the unit cell, with the boron (B) atoms occupying sites $(0,0,\frac{1}{2})$ and $(\frac{1}{3},\frac{2}{3},0)$ while the nitrogen (N) atoms are at $(0,0,0)$ and $(\frac{1}{3},\frac{2}{3},\frac{1}{2})$. Our self-consistent calculations were performed with the experimental lattices constants a=2.504 Å and c= 6.661Å, at room temperature. We expanded the radial parts of the orbitals in terms of even-tempered Gaussian functions. The s and p orbitals for the cation $B^{3+}$ were each described with 16 even-tempered Gaussian functions with the respective minimum and maximum exponents of 0.2658 and 0.1655 x $10^5$ for the atomic potential and 0.1242 and 0.1365 x $10^5$ for the atomic wave functions. The self-consistent calculations for $B^{3+}$ led to the total charge of 2.0005, which is also the valence charge, with an error per electron of 2.5 x $10^{-4}$. Similarly, the s and p orbitals for $N^{3-}$ were described with 20 even-tempered Gaussian functions with the respective minimum and maximum exponents of 0.1600 and 0.1600 x $10^5$ for the atomic potential and 0.1000 and 0.1300 x $10^5$ for the atomic wave functions. These exponents led to the convergence of the atomic calculations for $N^{3-}$ with the total, core and valence charges of 10.00004, 2.00002, and 8.00002, respectively. The error per electron was therefore 4 x $10^{-6}$. We utilized a 24 k-point mesh with proper weights, in the irreducible Brillouin zone, for the self-consistency iterations. The criterion for the convergence of the iterations was a difference of $10^{-5}$ or less between the potentials from two consecutive ones. We used 140 k points in the irreducible Brillouin zone for the production of the final, self-consistent bands.

3. Results

Table 3 below contains information on the successive calculations performed with the purpose of reaching the absolute minima of the occupied energies. The band gap generally can decrease or increase before one reaches the *optimal basis set*. As shown farther below, with the graphs of the bands, Calculations IV, V, and VI led to the same occupied energies indicating that these energies have reached their absolute minima, i.e., the ground state. As per the BZW-EF method, Calculation IV, the first of the three (3) is the one providing the DFT description of the material. The basis set for this calculation is the *optimal basis set*, i.e., the smallest basis set leading to the ground state of the material, without being over-complete.

**Table 3.** Successive calculations with the BZW-EF method, for h-BN (Calculations I-VI). In these calculations, the lattice constants are a = 2.504Å and c = 6.661Å, at room temperature. Calculation IV led to the absolute minimum of the occupied energies, given that Calculations V and VI produced occupied energies identical to corresponding ones from Calculation IV. The calculated indirect band gap, from near K to M, is 4.369 eV (or 4.37 eV).

| Calculation No. | Valence Orbitals for $B^{3+}$ | Valence Orbitals for $N^{3-}$ | No. of functions | Band gaps(eV) (near K-M) |
|---|---|---|---|---|
| I | $1s^2 2p^0 2s^0$ | $2s^2 2p^6$ | 36 | 7.499 |
| II | $1s^2 2p^0 2s^0$ | $2s^2 2p^6 3p^0$ | 48 | 5.767 |
| III | $1s^2 2p^0 2s^0 3p^0$ | $2s^2 2p^6 3p^0$ | 60 | 4.370 |
| **IV** | **$1s^2 2p^0 2s^0 3p^0$** | **$2s^2 2^6 3p^0 3s^0$** | **64** | **4.369** |
| V | $1s^2 2p^0 2s^0 3p^0 3s^0$ | $2s^2 2^6 3p^0 3s^0$ | 68 | 4.365 |

| VI | $1s^2 2p^0 2s^0 3p^0 3s^0$ | $2s^2 2p^6 3p^0 3s^0 4p^0$ | 80 | 4.210 |

Figures 1a through 1e below provide a graphical illustration of the generalized minimization of the energy, as the basis set is methodically augmented for successive, self-consistent calculations. Every pair of bands from consecutive calculations is shown below. In Fig. 1c, Calculations III may appear to reach the minima of the occupied energies, given that these occupied energies are mostly the same as corresponding ones from Calculation IV. However, a close examination of the occupied energies around -18.50 eV, at the Γ point, shows that both bands have been lowered by Calculation IV from their values from Calculation III. The occupied energies from Calculation IV are identical to the corresponding ones from Calculations V and VI. This perfect superposition of the occupied energies from three (3) consecutive calculations is the robust criterion for the attainment of the absolute minima of the occupied energies, i.e., the ground state of the material. As such, these occupied energies possess the full, physical content of DFT. From Figures 1d and 1e, it is apparent that the referenced superposition of the occupied energies does not hold for the all the unoccupied ones. It is instructive to note, however, that the low laying, unoccupied energies from the three (3) calculations, up to 8 eV, are also superimposed. This gratifying feature, notwithstanding, it is clear from the graphs that higher, unoccupied energies tend to be lowered as the size of the basis set increases.

The top of the valence band (VBM) is between K and Γ, at about 10% of the K-Γ separation, to the left of K. Its distance from K is $\Delta K = (4\pi/3a) \times 0.1 = 0.0885$ a.u, where $a = 4.7319$ a.u. is a lattice constant in atomic units. Hence, the location of the VBM is at K* = K-ΔK = (0, 0.7965, 0), to the left of K.

**Figures 1a – 1e.** Energy bands of hexagonal BN (h-BN) as obtained in Calculations I-VI of the BZW-EF method. These figures show the bands for pairs of consecutive calculations, with solid lines for bands of a calculation and dashed lines for the bands of the calculation immediately following it. The progressive lowering of the occupied energies, upon setting the Fermi levels to zero, is apparent, up to Calculation IV-VI, which produced the same absolute minima of the occupied energies, i.e., the ground state.

**Figure 1a** 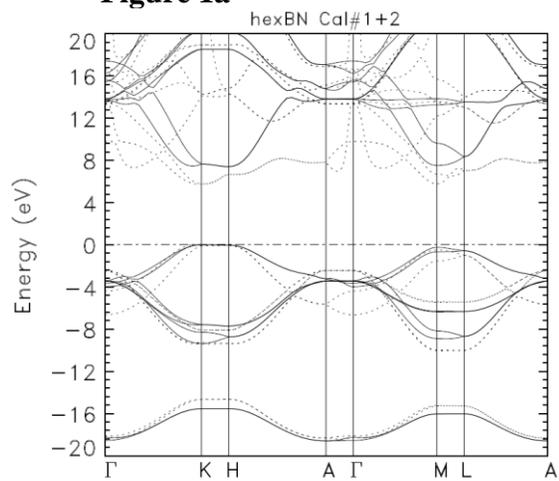

**Figure 1b** 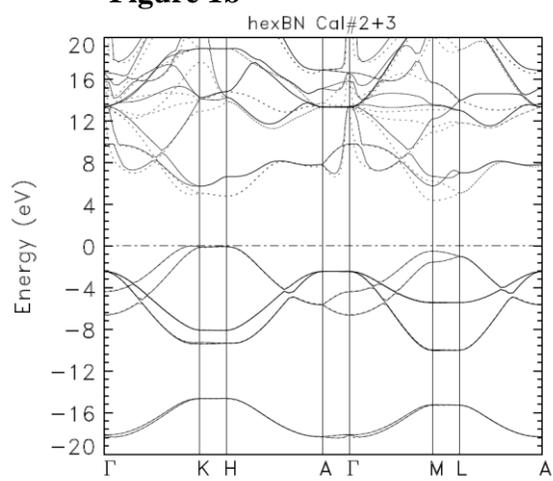

**Figure 1c**

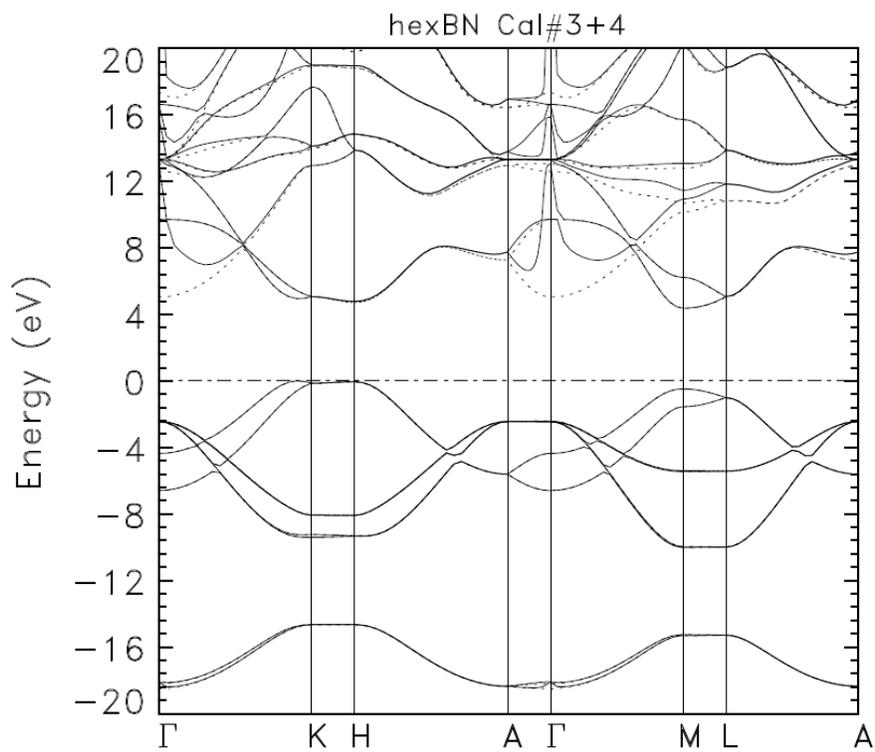

| Figure 1d | Figure 1e |
|---|---|
| 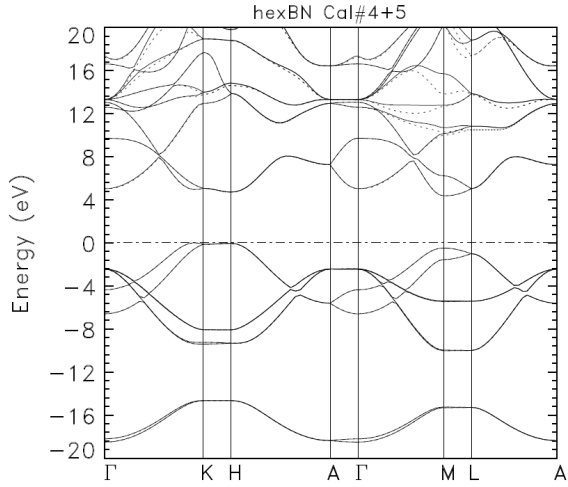 | 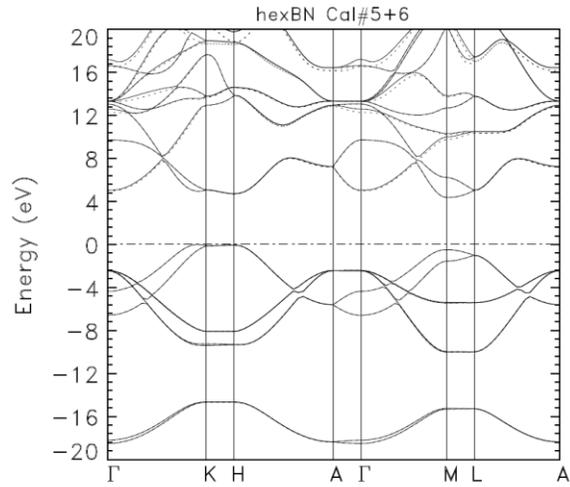 |

**Table 4**. Calculated, electronic energies (in eV) of h-BN, at high symmetry points in the Brillouin zone, obtained from Calculation IV. The Fermi energy is set equal to zero. The calculated band gap is 4.37 eV.

| Γ-point | K-ΔK-point | K-point | H-point | A-point | M-point | L-point |
|---|---|---|---|---|---|---|
| 17.357 | 21.116 | 20.759 | 19.791 | 16.4508 | 21.593 | 21.745 |
| 16.617 | 20.969 | 20.759 | 19.791 | 16.4508 | 21.259 | 21.745 |
| 13.322 | 19.793 | 18.939 | 18.802 | 13.320 | 21.033 | 18.820 |
| 13.321 | 18.613 | 18.939 | 18.802 | 13.320 | 20.236 | 18.820 |
| 13.305 | 16.896 | 17.668 | 14.843 | 13.319 | 15.699 | 13.856 |
| 13.304 | 14.445 | 13.994 | 14.843 | 13.319 | 12.780 | 13.856 |
| 13.056 | 13.656 | 13.994 | 13.878 | 12.958 | 10.689 | 10.824 |
| 12.592 | 12.309 | 12.957 | 13.878 | 12.958 | 10.163 | 10.824 |
| 9.714 | 5.445 | 5.064 | 4.715 | 7.263 | 6.222 | 5.040 |
| 5.049 | 4.953 | 5.064 | 4.715 | 7.263 | **4.369** | 5.040 |
| -2.419 | **0.000** | -0.138 | **-0.048** | -2.435 | -0.482 | -1.007 |
| -2.420 | -0.614 | -0.138 | -0.048 | -2.435 | -1.552 | -1.007 |
| -2.453 | -7.827 | -8.067 | -8.082 | -2.436 | -5.399 | -5.423 |
| -2.453 | -7.833 | -8.067 | -8.082 | -2.436 | -5.452 | -5.423 |
| -4.365 | -9.241 | -9.242 | -9.322 | -5.606 | -9.960 | -9.990 |
| -6.593 | -9.366 | -9.400 | -9.322 | -5.606 | -10.012 | -9.990 |
| -18.206 | -14.748 | -14.653 | -14.653 | -18.368 | -15.254 | -15.283 |
| -18.509 | -14.801 | -14.653 | -14.653 | -18.368 | -15.313 | -15.283 |

Even though the occupied energies in Table 4 and the graph of the bands from Calculation IV (in Figure 1d) provide an adequate description of the ground state electronic properties of hexagonal BN, we discuss farther below subtilities relative to the valence band maximum (VBM) and the conduction band minimum (CBM). In particular, our close examination of the bands hints at a possible explanation of the multitude of VBM-CBM pairs reported by previous density functional theory calculations. These calculations, as far as we can determine, did not performed the generalized minimization of the energy as dictated by the second DFT theorem,

Figures 2 and 3 respectively show the calculated, total and partial densities of states (DOS, pDOS). We derived them from the bands produced by Calculation IV, with the optimal basis set. Short, vertical segments indicate the locations of major peaks, whose values are provided on the graph of the total density of states. The calculated valence band width of 18.58 eV is in agreement with the calculated valence band width (18.5 eV) from Ma et al. [23] and from Castellani et al. [24]. While this value is smaller than the experimental finding of 20.7 $\pm$ 1.5 eV obtained by J. Barth et al. [7] and by Tegeler et al. [8] in their XPS measurements, we note that, according to these authors [7, 8, 23, 24], the real total width of the valence bands may be smaller than the measured value by l-3 eV, due to significant Auger broadening of the XPS spectrum at energies corresponding to the s band.

The lower and upper groups of valence bands have widths of 3.98 eV and 10.02 eV, respectively. Three major peaks in the density of states for the conduction bands are located at 4.92 eV, 12.88 eV, and 18.46 eV. The above characteristics of the total density of states (DOS), for h-BN, will be hopefully confirmed by future experimental measurements. Additionally, the eigenvalues in Table 4 lend themselves to comparison with some X-Ray and UV spectroscopic measurements. From Fig.3, for the partial densities of state (pDOS), we clearly observe a net dominance by nitrogen s state in the lowest group of valence bands, with a tiny contribution from boron p state. In the upper group of valence bands, N p dominates, with small contributions from boron p and minuscule ones from boron s. This hybridization of nitrogen p and boron p should be observable in X-Ray spectroscopic measurements. While the largest contribution to the conduction bands comes from nitrogen p, particularly around the absorption edge, that of boron p is also significant. Both N s and B s have evanescent contributions to the conduction bands.

**Figure 2**. Calculated, total density of states (DOS) for hexagonal boron nitride (h-BN), obtained with the bands from Calculation IV.

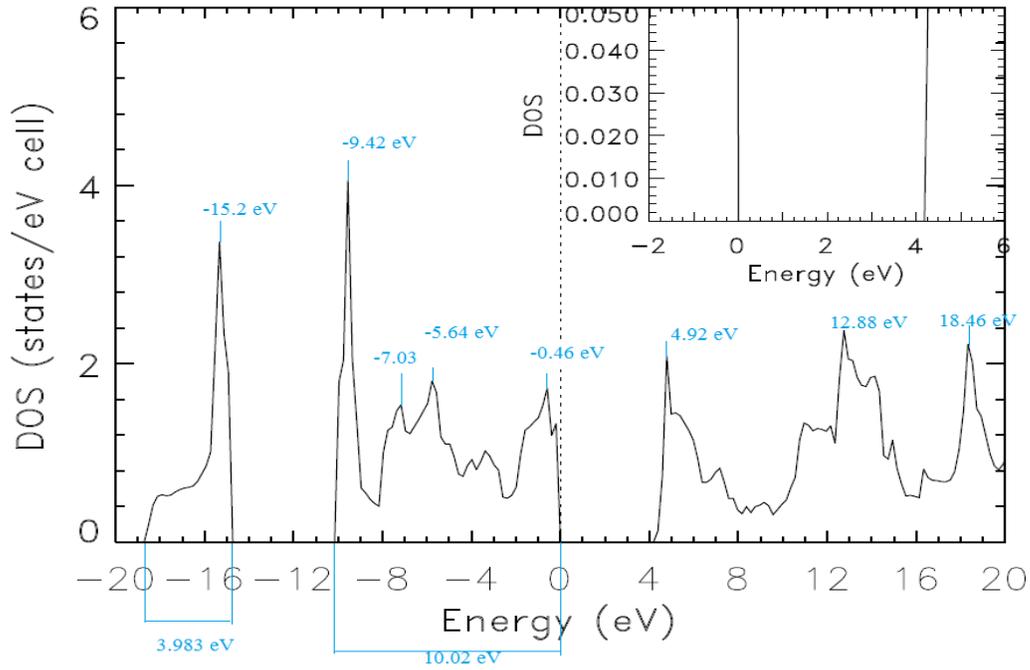

**Figure 3**. Calculated, partial densities of states (p-DOS), as derived from bands resulting from Calculation IV.

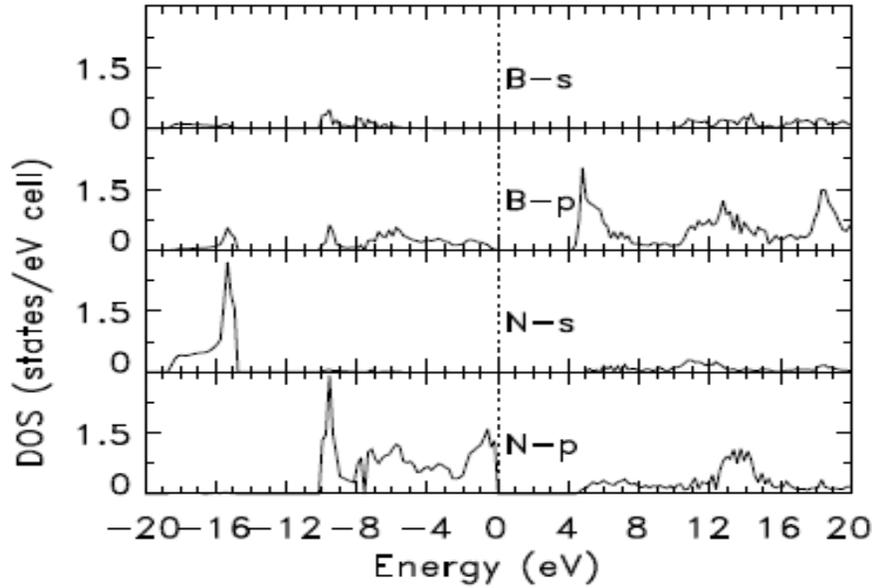

Several transport properties, including various mobilities for electrons or holes, depend on the inverse of the electron or hole effective masses, respectively. For this reason, we have calculated the electron and hole effective masses shown below in Table 5, in units of the electron mass $m_0$. With values of 0.205 $m_0$, 2.250 $m_0$, and 1.730 $m_0$ in the M to Γ, M to K, and M to L directions, respectively, the electron effective mass at the bottom of the conduction band is clearly anisotropic. The same is true for the electron effective mass at H, even though its values from H to A and H to Γ are identical.

The hole effective masses from K* to Γ, K* to H, and K* to M are respectively 0.534, 0.569, and 1.48, in units of $m_0$. The calculated hole effective masses at the H symmetry point, along H-A, H-Γ, H-K, and H-L axes, are 0.822, 0.822, 3.468, and 1.671, respectively, in units of $m_0$. These hole effective masses are anisotropic, despite the equality of the ones from H to A and H to Γ.

**Table 5**. Calculated effective masses for hexagonal BN, in units of free electron mass $m_0$: $M_e$ indicates an electron effective mass in the conduction bands and $M_h$ represents a hole effective mass. The top of the valence band is at K*, to the left of the K symmetry point, as defined above.

| Types and Directions of Effective Masses | Values of Effective Masses ($m_o$) |
|---|:---:|
| $M_e$(M-Γ) | 0.205 |
| $M_e$(M-K) | 2.250 |
| $M_e$(M-L) | 1.730 |
| $M_e$(H-A) | 0.588 |
| $M_e$(H-Γ) | 0.588 |
| $M_e$(H-K) | 1.102 |
| $M_e$(H-L) | 3.129 |
| $M_e$(K-Γ) | 0.387 |
| $M_e$(K-H) | 0.433 |
| $M_h$(K*-Γ) | 0.534 |
| $M_h$(K*-H) | 0.569 |
| $M_h$(K*-M) | 1.480 |
| $M_h$(H-A) | 0.822 |
| $M_h$(H-Γ) | 0.822 |
| $M_h$(H-K) | 3.468 |
| $M_h$(H-L) | 1.671 |

## 4. Discussion

A discussion of our results, particularly in relation to findings from previous DFT calculations, rests on the following fact. None of the previous calculations appear to have performed a generalized minimization of the energy. The minimization obtained following self-consistent iterations, with a single basis set, produces the minimum of the energy relative to that basis set. Such solutions are stationary ones whose number is practically infinite. None should be *à priori* assumed to provide a description of the ground state of the material. Consequently, the computational results should not be expected to possess the full, physical content of DFT or to agree with experimental measurement. Our generalized minimization, as thoroughly explained above, verifiably leads to the absolute minima of the occupied energies, i.e., the ground state, as required by the second DFT theorem. Explicitly searching for the ground state and avoiding basis sets that are overcomplete for the description of the ground state are two requirements for a correctly performed DFT calculation. We address below plausible, negative consequences use of over-complete basis sets.

With the second corollary of the first DFT theorem, i.e., that the spectrum of the Hamiltonian is a unique functional of the ground state charge density [34], we avoid over-complete basis sets. While these larger basis sets lead to the ground state energies, they also lower some unoccupied energies from their values obtained with the *optimal basis set*. As per the above corollary, any unoccupied energy, different from (i.e., lower than) its corresponding value obtained with optimal basis set, no longer belongs to the spectrum of the Hamiltonian. This rigorous conclusion also results from the fact that, with these larger basis sets, the charge density and the Hamiltonian do not change from their respective values obtained with the optimal basis set. Consequently, the unoccupied eigenvalues, different from their corresponding values obtained with the *optimal basis set*, cannot rationally be physically meaningful ones. The Rayleigh theorem for eigenvalues, as elaborated upon elsewhere [34, 49-50], trivially explains the spurious lowering of unoccupied energies in calculations employing larger basis sets that contain the *optimal one*. We should note the spuriously lowered, unoccupied energies, including some lowest laying ones, provide one plausible explanation of the widespread underestimation of band gaps in the literature. This contention stems in part from the fact that single basis set calculations tend to employ large basis sets in order to avoid incompleteness.

With the above understanding, we discuss the fine structures of the bands using the enlarged graphs in Figures 4 and 5 below. While Figure 4 shows the entire band structure, Figure 5 only exhibits the drastically enlarged uppermost and lowest valence and conduction bands, respectively, around and between the K and H symmetry points. In Figure 4, the highest and degenerate valence bands are visibly close to the Fermi level, from K to H. Figure 5 is needed to ascertain the location of the valence band maximum. To do so, one is guided by the fact that, at the location of the VBM, the band is superimposed on a short segment at the Fermi level. Figure 5 shows that the VBM is at the K* point defined above. At the H point, the degenerated valence band is only 0.048eV below the Fermi level. The top of the valence band at M is 0.482 eV below the Fermi level. The direct band gap at M is therefore 4.369 eV + 0.482 eV = 4.851 eV. It is slightly larger than the one at H which is 4.763 eV + 0.048 eV = 4.811 eV.

**Figure 4**. The enlarged graph of the band structure of hexagonal BN, produced by Calculation IV, with the optimal basis set.

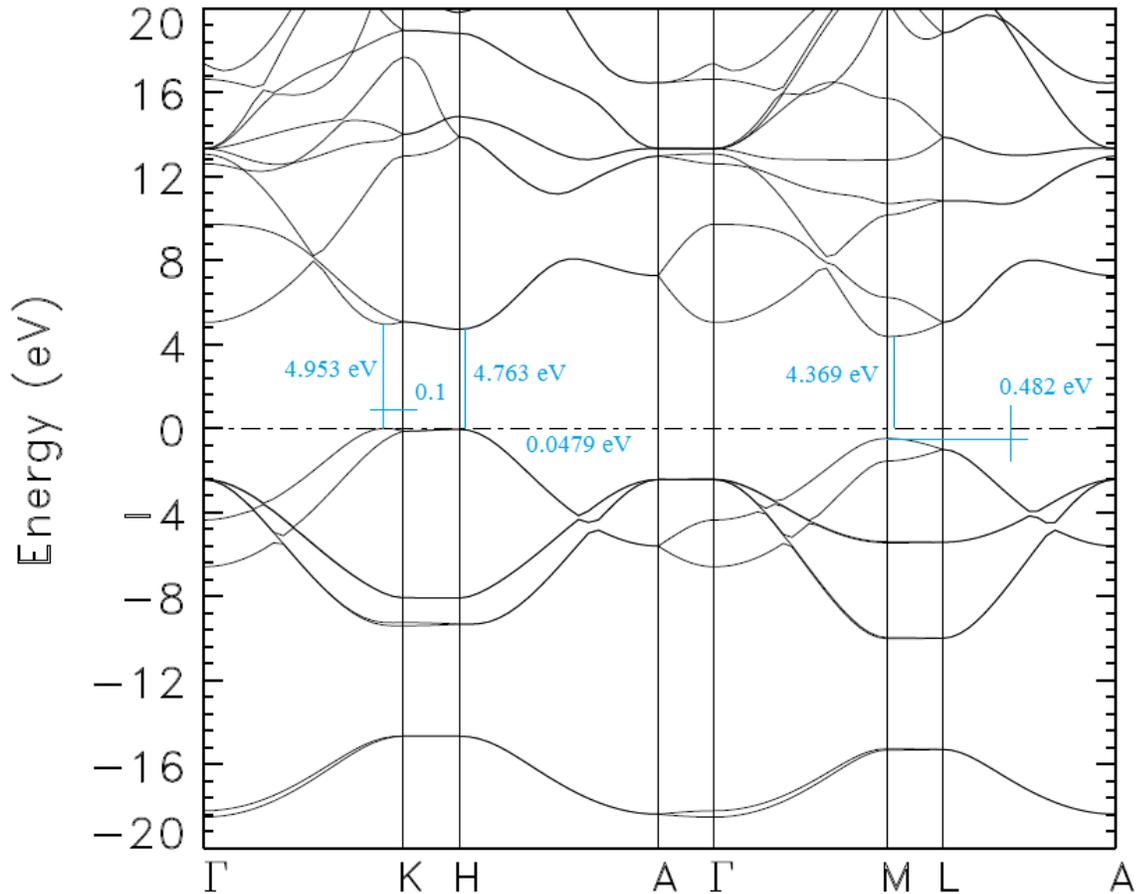

**Figure 5**. The further enlarged parts of highest and lowest valence and conduction bands, respectively, in Figure 4, between and around the K and H high symmetry points. Clearly, the top of the valence band is the only part that is superimposed on the Fermi level; this top is at K* as defined above, to the left of K.

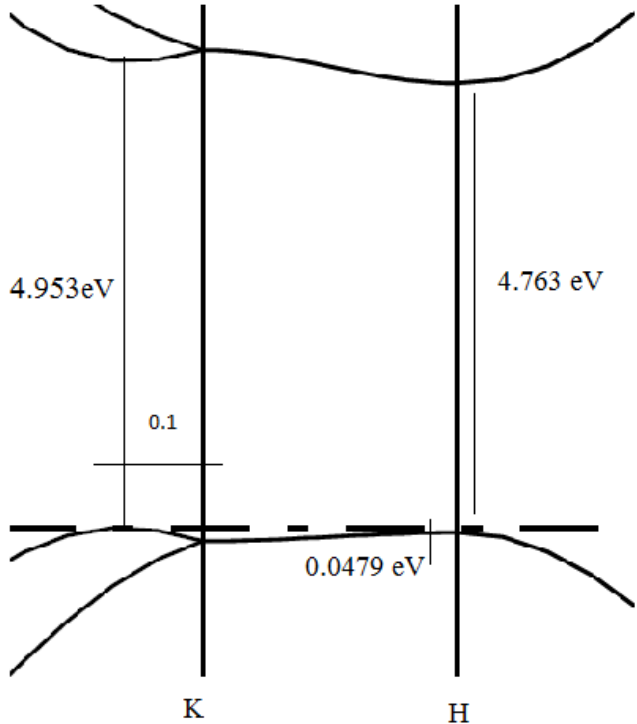

The above fine structures of the bands hint to a possible explanation of the report of seven (7) different VBM-CBM pairs by previous DFT calculations. Indeed, while the presumed single basis sets in these calculations may be close to or contain the corresponding optimal basis sets, with the above subtle features of the band structure, the slightest deviation of these basis sets from the one describing the ground state could explain the differences between the resulting bands and between them and the ones reported here. Additionally, without the generalized minimization, it is practically hopeless to have the basis set complete for the description of the ground state, without being over-complete.

## 5. Conclusion

We have presented the description of electronic and related properties of the ground state of h-BN, as obtained from ab-initio, self-consistent density functional theory (DFT) calculations. Our generalized minimization of the energy, following the BZW-EF method, verifiably led to the ground state and avoided over-complete basis sets. Our findings possess the full, physical content of DFT. Our calculated indirect band gap from K* to M is 4.37 eV. This value is practically in agreement with the experimental finding of 4.30 eV which is the most accepted one in the literature. The density of states (DOS) and partial densities of states (p-DOS) are in good agreement with those from electron momentum spectroscopy (EMS) [6-9]. To the best of our knowledge, no measurements of the electron effective masses are available for comparison with our calculated ones. In light of our previous success, partly through accurate predictive capabilities, we expect future experiments to confirm our findings.

**Acknowledgments**:

This work was funded in part by the National Science Foundation [NSF, Award Nos. EPS-1003897, NSF (2010-2015)-RH SUBR, and HRD-1002541], the US Department of Energy, National Nuclear Security Administration (NNSA, Award No. DE-NA0002630), LaSPACE and LONI-SUBR.